\documentclass[12pt]{article}
\usepackage{graphicx}
\usepackage{multirow}
\usepackage{subfig}
\usepackage{hyperref}
\usepackage{lineno}

\newcommand\pubnumber{CIPANP2018-Reynolds}
\newcommand\pubdate{\today}

\def\Title#1{\begin{center} {\Large #1 } \end{center}}
\def\Author#1{\begin{center}{ \sc #1} \end{center}}
\def\Address#1{\begin{center}{ \it #1} \end{center}}

\newcommand\pubblock{\rightline{\begin{tabular}{l} \pubnumber\\
         \pubdate  \end{tabular}}}
\newenvironment{Abstract}{\begin{quotation}  }{\end{quotation}}
\newenvironment{Presented}{\begin{quotation} \begin{center} 
             PRESENTED AT\end{center}\bigskip 
      \begin{center}\begin{large}}{\end{large}\end{center} \end{quotation}}

\def\beq{\begin{equation}}
\def\eeq#1{\label{#1}\end{equation}}
\def\eeqn{\end{equation}}

\def\beqa{\begin{eqnarray}}
\def\eeqa#1{\label{#1}\end{eqnarray}}
\def\eeqan{\end{eqnarray}}

\let\bar=\overbar

\def\Dslash{\not{\hbox{\kern-4pt $D$}}}
\def\dslash{\not{\hbox{\kern-2pt $\del$}}}

\def\msb{{\bar{\ssstyle M \kern -1pt S}}}

\begin{document}

\begin{titlepage}
\pubblock

\vfill
\Title{Searches for rare and non-Standard Model decays of the Higgs boson}
\vfill
\Author{Elliot Reynolds$^*$, on behalf of the ATLAS Collaboration}
\Address{$^*$University of Birmingham}
\vfill
\begin{Abstract}
  Discovered in 2012, the Higgs boson has opened a new window on nature. The latest searches for its rare and non-Standard Model decays with the ATLAS detector at the LHC are presented. They represent a promising probe of the 1st and 2nd generation Yukawa couplings, and of new physics. Searches for rare exclusive decays of the Higgs boson to a meson and a photon are presented. Also presented are four searches for decays of the Higgs boson to pairs of beyond the Standard Model resonances, in various final states.
\end{Abstract}
\vfill
\begin{Presented}
  CIPANP 2018 - Conference on the Intersections of Particle and Nuclear Physics\\
  Palm Springs, CA, USA, 29 May -- 3 June, 2018
\end{Presented}
\vfill
\footnotesize\textcopyright 2018 CERN for the benefit of the ATLAS Collaboration.\\
\footnotesize Reproduction of this article or parts of it is allowed as specified in the CC-BY-4.0 license.
\end{titlepage}

\section{Introduction}

As the only seemingly fundamental scalar in the Standard Model (``SM''), associated with an all-permeating, ever-present field which generates the mass of all SM particles, the discovery of the Higgs boson has opened a new window through which nature can be probed. While initial studies into its properties are compatible with the SM predictions~\cite{couplings,masscouplingscms,masscouplingszcms,spin0,anomolouscouplingscms,couplingsatlasandcms,massatlasandcms}, further studies are required to establish a full understanding of the nature of the Higgs sector. Two such avenues are searches for rare decays and beyond the SM (``BSM'') decays of the Higgs boson, with the ATLAS detector~\cite{atlas} at the LHC~\cite{lhc}.

These searches share two common themes. First, these are at best rare, or potentially non-existent processes. No significant excesses have yet been observed. As such $95\%$ confidence level upper limits (``limits'') are set according to the $CL_s$ prescription~\cite{CLs1,CLs2}, using maximum likelihood fits (``fits''), sometimes to multiple variables simultaneously. Second, due to the low expected yield of these processes, the dominant uncertainty for most of the analyses shown is due to limited data statistics.

\section{Rare Decays}


In the SM, six of the twenty-six fundamental constants of nature are the Yukawa couplings of the first and second generation fermions. To complete our understanding of nature and access these constants, measuring the rare decays of the Higgs boson is necessary. Their small SM cross section also make these decay modes potentially susceptible to notable modifications from new physics.



Many searches for rare decays of the Higgs boson have been performed at the ATLAS detector, most of which can not be covered here. For example, see Refs~\cite{Hcc,Hmumu,Hinv,HZgam}. The following subsection describes the subset of these searches for which the Higgs boson decays to a meson ($M$) and a photon ($\gamma$).

\subsection{\boldmath $H\to M\gamma$}

\begin{figure}[htb]
  \centering
  \subfloat[Direct]{\includegraphics[width=0.275\textwidth]{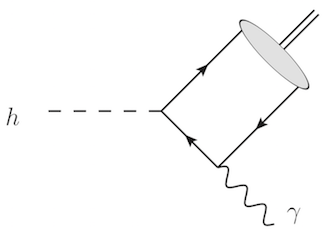}}
  \hspace{0.15\textwidth}
  \subfloat[Indirect]{\includegraphics[width=0.275\textwidth]{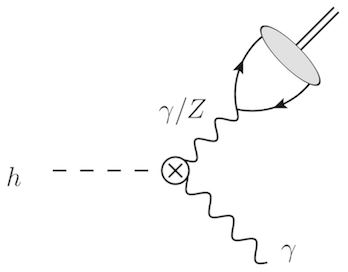}}
  \caption{Direct and indirect decays of the Higgs boson to a meson and a photon~\cite{HMgamCalc1}.}
  \label{fig:hmgammodes}
\end{figure}

Rare exclusive decays of the Higgs boson to a meson and a photon provide access to the first and second generation Yukawa couplings, while providing a final state with a distinct topology to trigger and select events~\cite{hmgam1,hmgam2,hmgam3}. These final states can be produced directly or indirectly as described by the Feynman diagrams shown in Figure~\ref{fig:hmgammodes}. Only the direct diagram gives access to the Yukawa coupling, unfortunately it is subdominant and the diagrams interfere destructively. The mesons targeted are listed in Table~\ref{tab:hmgamfinalstates}.

\begin{table}[t]
  \begin{center}
    \begin{tabular}{|l|l|l|}
      \hline
      Target Meson & $BR_{\mathrm{SM}}(H\rightarrow M\gamma)$ & Meson Decay Mode\\
      \hline
      $\rho$ & $1.7\times 10^{-5}$ & $\rho\rightarrow\pi^+\pi^-$\\
      $\phi$ & $2.3\times 10^{-6}$ & $\phi\rightarrow K^+ K^-$\\
      $J/\psi$ & $2.8\times 10^{-6}$ & $J/\psi\rightarrow\mu^+\mu^-$\\
      $\Upsilon(1S,2S,3S)$ & $(6.1,2.0,2.4)\times 10^{-10}$ & $\Upsilon\rightarrow\mu^+\mu^-$\\
      \hline
    \end{tabular}
    \caption{List of searches with final states including a photon and a meson. Also shown are the expected branching ratio of the Higgs boson to these final states in the SM~\cite{HMgamCalc1,HMgamCalc2}, and the targeted decay mode of the meson.}
    \label{tab:hmgamfinalstates}
  \end{center}
\end{table}

\begin{figure}[htb]
  \centering
  \subfloat[$\rho\rightarrow\pi^+\pi^-$]{\includegraphics[width=0.45\textwidth]{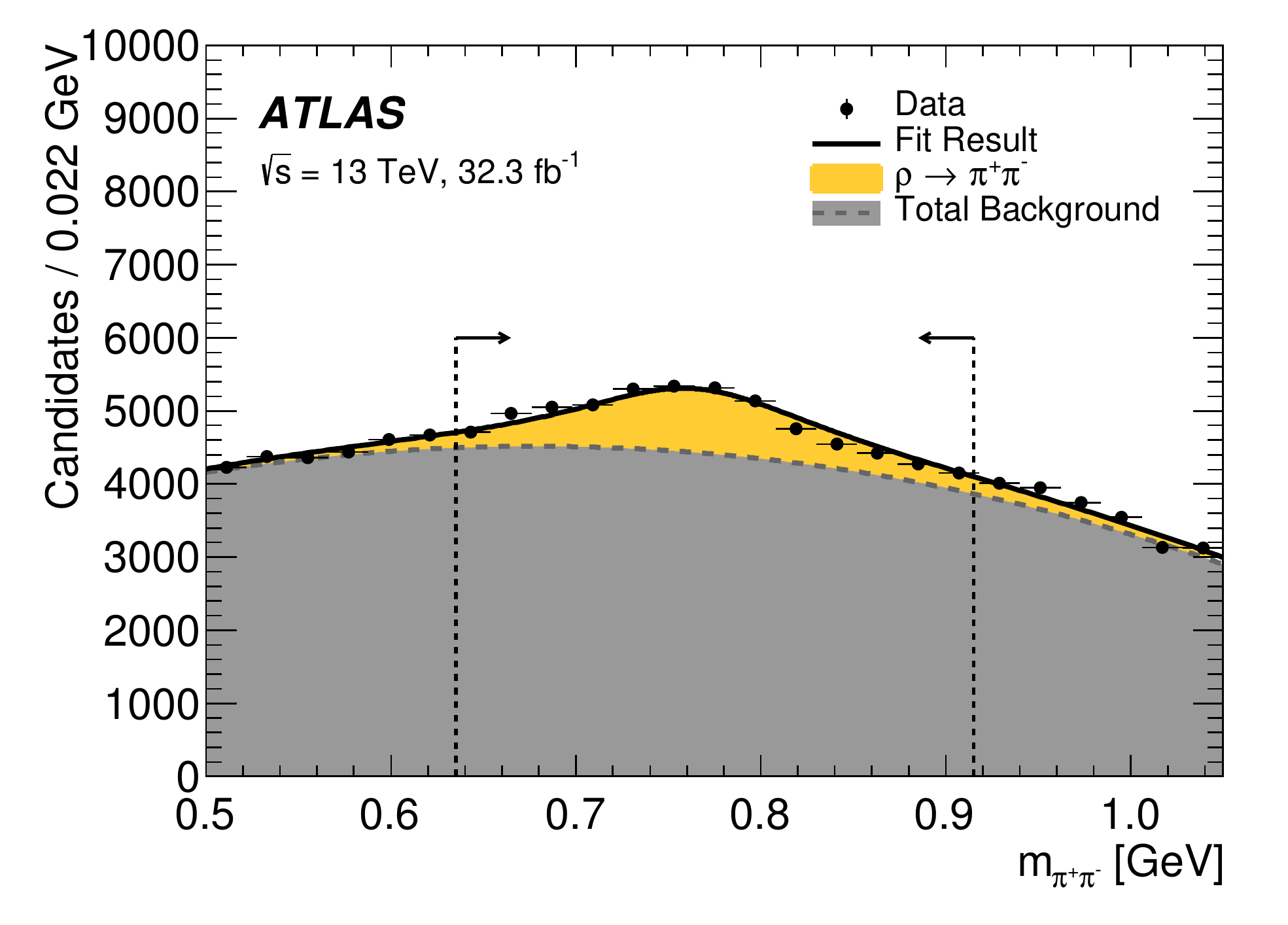}}
  \hspace{0.033333333333\textwidth}
  \subfloat[$\phi\rightarrow K^+ K^-$]{\includegraphics[width=0.45\textwidth]{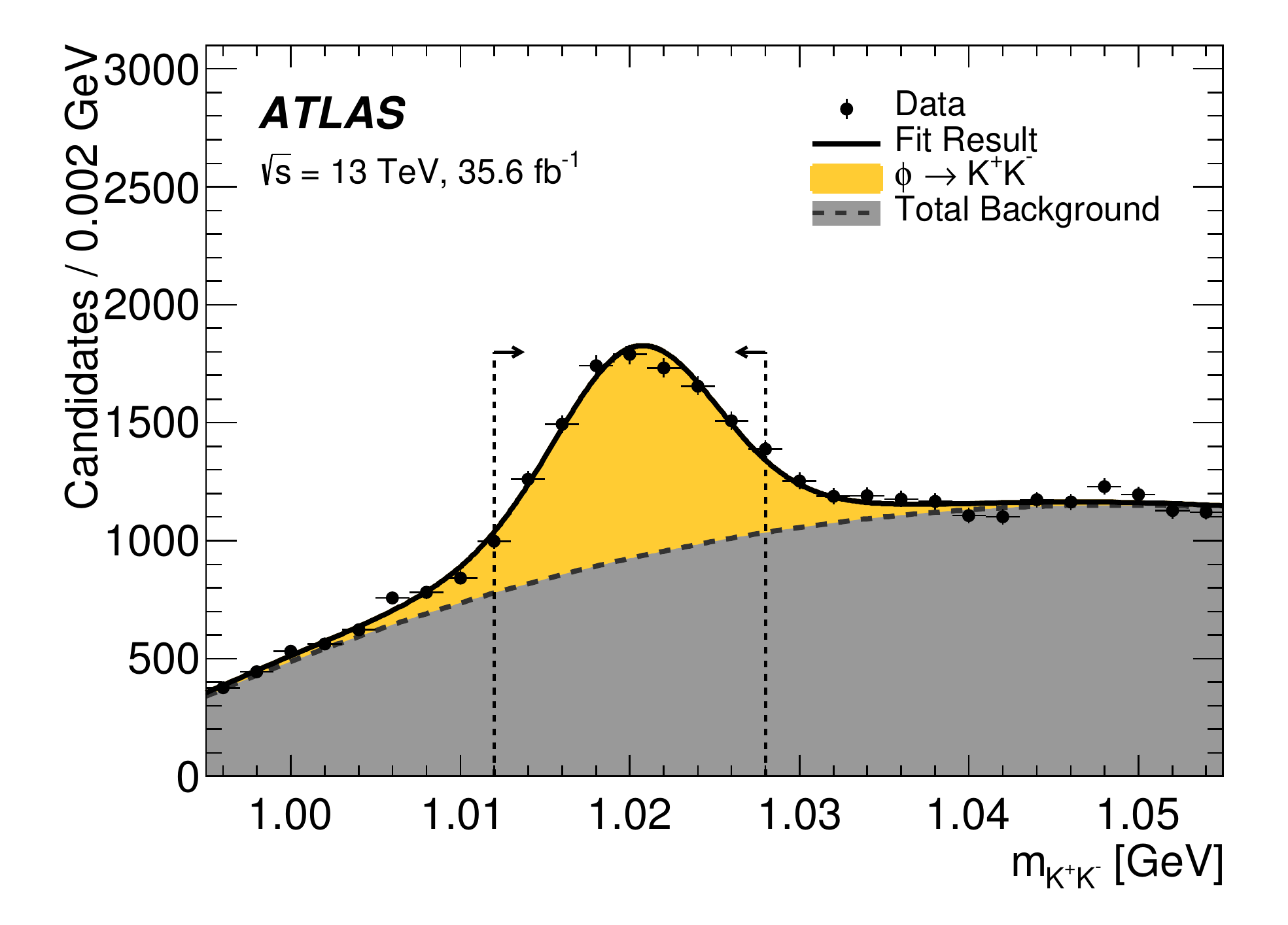}}
  \caption{Invariant masses for the di-track systems resulting from the meson decay, in the generation regions of the $H\rightarrow \rho\gamma$ (a) and $H\rightarrow\phi\gamma$ (b) searches~\cite{hmgam1}.}
  \label{fig:hmgamditrackmass}
\end{figure}

Events are triggered in the $\rho$ ($\phi$) searches by the $\pi^\pm$ ($K^\pm$) tracks and photon, while the $J/\psi$ and $\Upsilon$ searches use the muons to trigger. Jet-plus-photon production and di-jets are the dominant backgrounds for the $\rho$ and $\phi$ final states, while inclusive SM quarkonia production in association with a jet misidentified as a photon is the dominant background for the $J/\psi$ and $\Upsilon$ final states. In the $J/\psi$ and $\Upsilon$ searches, events are categorised based on whether or not the photon converted to an electron/positron pair in the inner tracker, and whether the muons are found in the barrel or end-caps of the detector. Background estimates are derived in all cases by using a non-parametric data-driven approach describing the kinematic distributions with templates, while the signal is modelled using either a Gaussian or double-Gaussian distribution in $m_{M\gamma}$. The selection applied for all final states includes a requirement on the invariant mass of the meson decay products, derived in a high statistics region. Figure~\ref{fig:hmgamditrackmass} shows the di-track invariant mass in such a region for the $\rho$ and $\phi$ searches.

\begin{figure}[htb]
  \centering
  \subfloat[$m_{\mu\mu\gamma}$]{\includegraphics[width=0.3\textwidth]{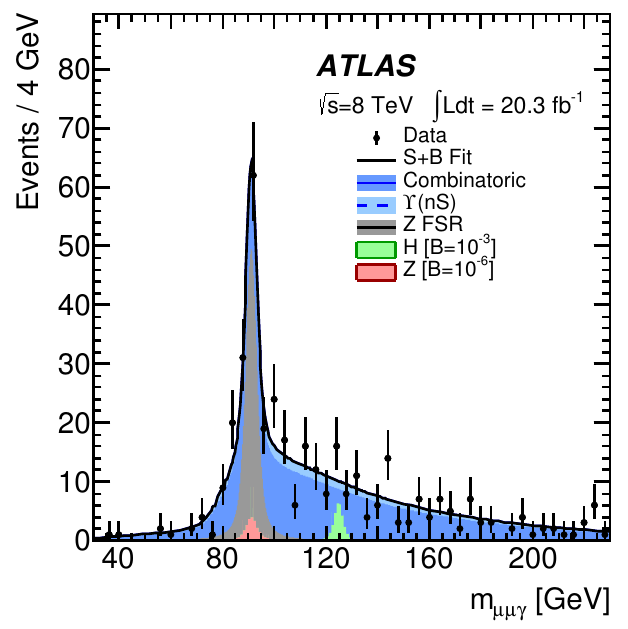}}
  \subfloat[$p_\mathrm{T}^{\mu\mu\gamma}$]{\includegraphics[width=0.3\textwidth]{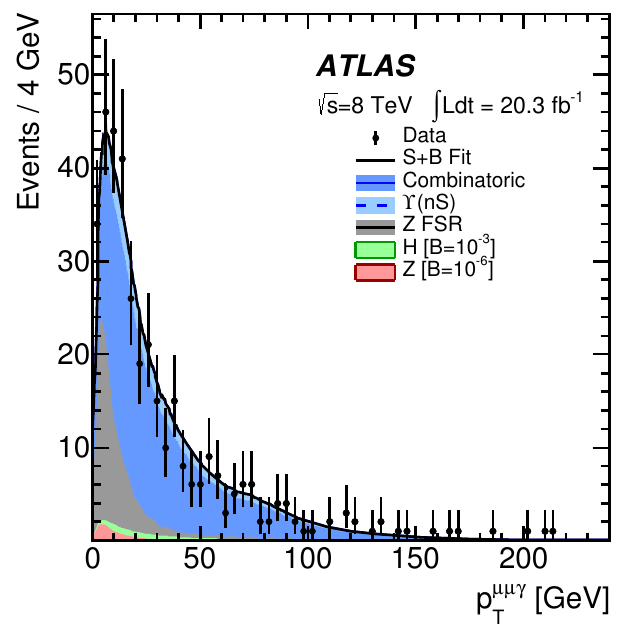}}
  \subfloat[$m_{\mu\mu}$]{\includegraphics[width=0.3\textwidth]{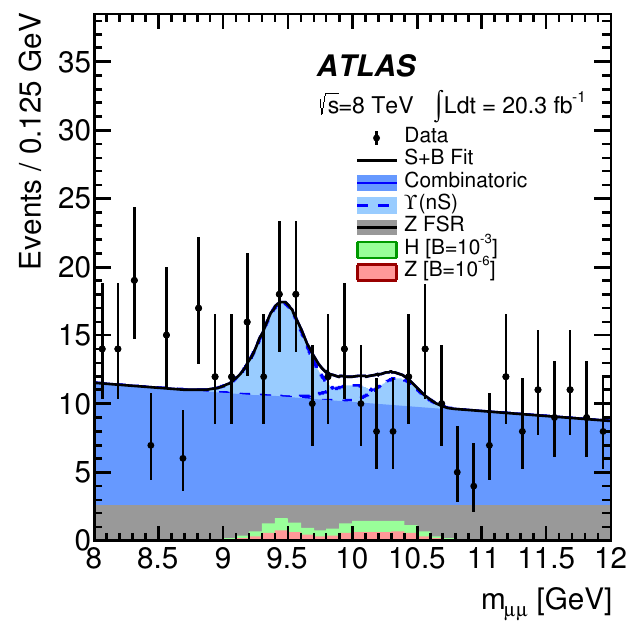}}
  \caption{$m_{\mu\mu\gamma}$, $p_\mathrm{T}^{\mu\mu\gamma}$ and $m_{\mu\mu}$ distributions, which were input to the fit of the $H\rightarrow \Upsilon\gamma$ search to set the limit, as an example of the $H\rightarrow M\gamma$ fit inputs~\cite{hmgam2}.}
  \label{fig:hmgamfitupsilon}
\end{figure}

All of these analyses use the $m_{M\gamma}$ distributions in the fits, while the $J/\psi$ and $\Upsilon$ analyses also include the $p_\mathrm{T}^{M\gamma}$ distributions, which provide additional discrimination. The $\Upsilon$ analysis also fits the $m_{M}$ distribution to distinguish the different $\Upsilon$ resonances. Figure~\ref{fig:hmgamfitupsilon} shows the distributions fitted in the $\Upsilon$ search. Limits were set of:

\begin{itemize}
\setlength\itemsep{0mm}
\item $\mathrm{BR}(H\rightarrow\rho\gamma)<8.8\times 10^{-4}$
\item $\mathrm{BR}(H\rightarrow\phi\gamma)<4.8\times 10^{-4}$
\item $\mathrm{BR}(H\rightarrow J/\psi\gamma)<1.5\times 10^{-3}$
\item $\mathrm{BR}(H\rightarrow\Upsilon(1S,2S,3S)\gamma)<(1.3,1.9,1.3)\times 10^{-6}$
\end{itemize}




\section{BSM Decays}





As collider energies increase, analyses often search for new higher mass states. However, new states could also exist at lower masses, undiscovered by previous experiments, if their only significant coupling is to the Higgs boson, which has only been produced in a sufficient abundance at the LHC. Therefore, the discovery of the Higgs boson unlocks a new range of searches for low mass particles predicted by various BSM models. For many of these models, the tightest constraint comes from fits to the major bosonic decay channels of the Higgs boson ($H\rightarrow ZZ\ \&\ \gamma\gamma$), which provide model-dependent indirect constraints on the branching ratios of $\sim$20\%~\cite{20pc}.

Two models considered are the two-Higgs doublet model (2HDM) and two-Higgs doublet model with an additional singlet state (2HDM+s), in which the Higgs boson can couple to light pseudoscalar Higgs resonances ($a^0$). These models are motivated because the Higgs sector need not have taken its simplest form, as it does in the SM. The Type-II 2HDM+s models~\cite{a0brs} also corresponds to the Higgs sector in a next to minimal version of Supersymmetry, which solves the $\mu$-problem of Supersymmetry~\cite{nmssm}, while greatly reducing the fine-tuning and little-hierarchy problems. Hidden dark-sector models, such as the Hidden Abelian Higgs Model~\cite{hahm}, also predict an additional resonance ($Z_d$), which can have a significant coupling to the Higgs boson.

The following subsections describe searches for decays of the observed Higgs boson to these resonances with masses between about 15 and 60 GeV (with one exception). For other such searches, see for example Refs~\cite{mumutautau,gamgamgamgam}. Below about 15 GeV, the final state particles leave overlapping showers in the calorimeters, for which no dedicated reconstruction exists in ATLAS. Above about 60 GeV, kinematic effects lower the branching ratio and background from SM vector bosons overwhelms the signal.

\subsection{\boldmath $H\rightarrow a^0a^0\ \&\ Z_dZ_d\rightarrow \ell\ell\ell\ell$}

\begin{figure}[htb]
  \centering
  \subfloat[$a^0a^0$]{\includegraphics[width=0.275\textwidth]{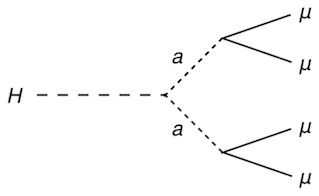}}
  \hspace{0.04375\textwidth}
  \subfloat[$Z_dZ_d$]{\includegraphics[width=0.275\textwidth]{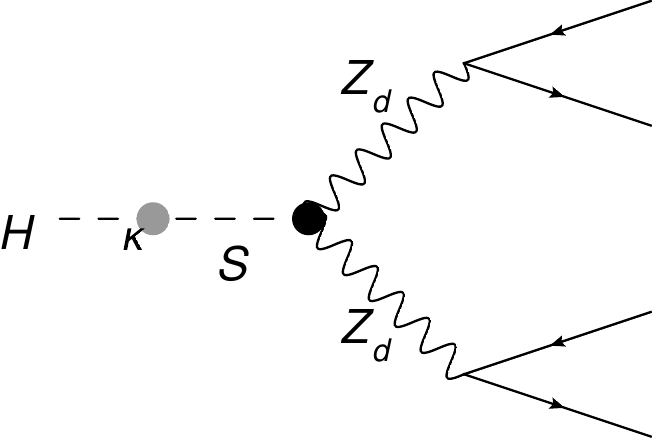}}
  \hspace{0.04375\textwidth}
  \subfloat[$ZZ_d$]{\includegraphics[width=0.275\textwidth]{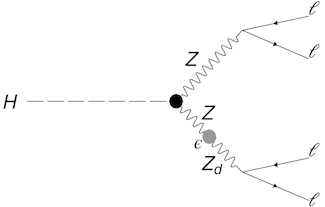}}
  \caption{Feynman diagrams for the $H\rightarrow a^0a^0\ \&\ Z_dZ_d\ \&\ ZZ_d\rightarrow \ell\ell\ell\ell$ processes~\cite{Leney:2282407,hxx4l}.}
  \label{fig:hxxproduction}
\end{figure}

\begin{figure}[htb]
  \centering
  \includegraphics[width=0.45\textwidth]{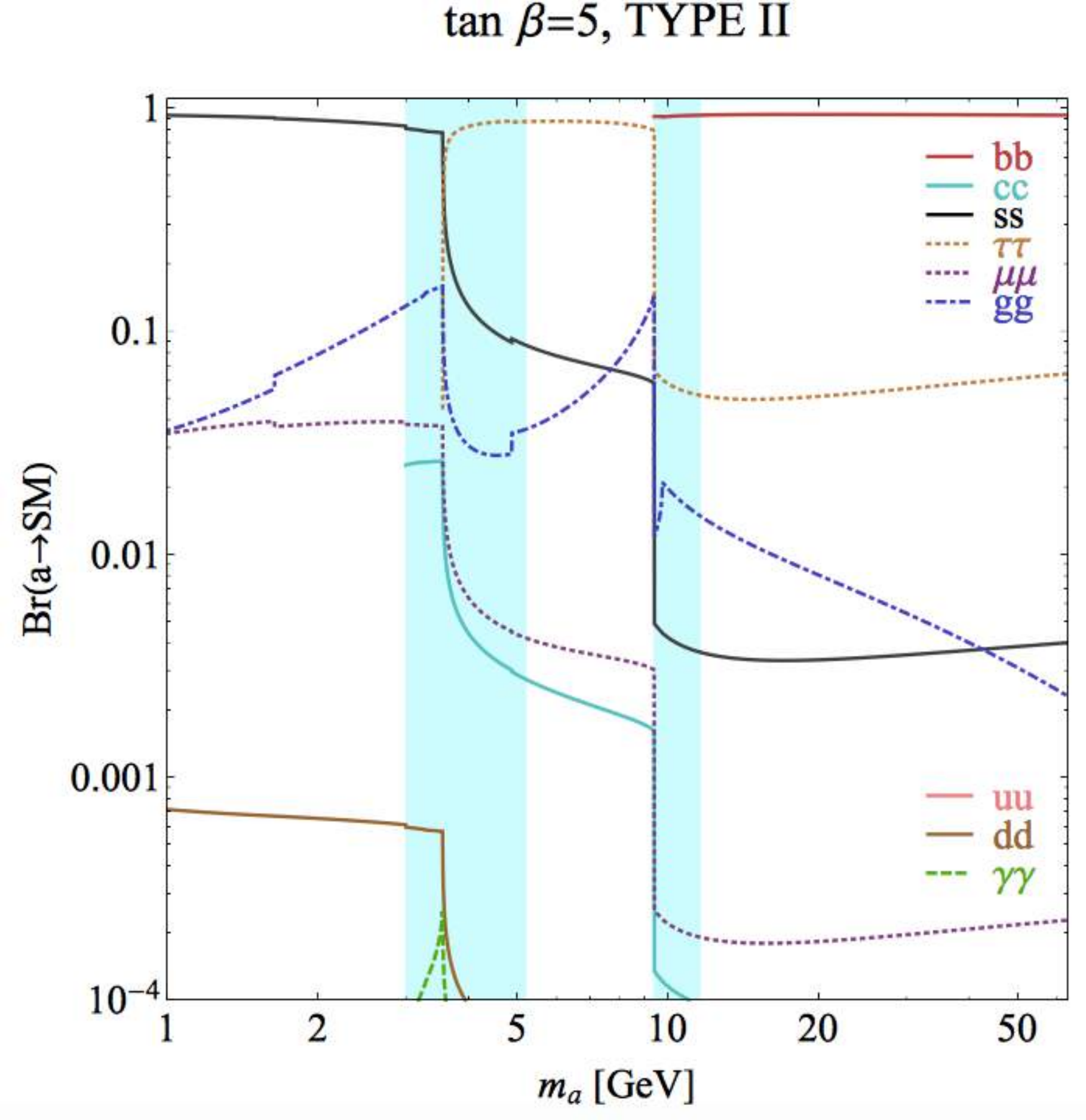}
  \caption{Branching ratios for the decay of the $a^0$ pseudoscalar Higgs resonance to SM particles, in the Type-II 2HDM, for $\tan\beta=5$~\cite{a0brs}.}
  \label{fig:a0brs}
\end{figure}

Leptons (electrons and muons) provide distinct signatures in the ATLAS detector, making $H\rightarrow a^0a^0\ \&\ Z_dZ_d\rightarrow \ell\ell\ell\ell$ a promising final state for searches for rare decays of the Higgs boson~\cite{hxx4l}. The Feynman diagrams for these processes are shown in Figure~\ref{fig:hxxproduction}(a,b), while the branching ratios are shown in Figure~\ref{fig:a0brs}. For the $a^0$ interpretation of this search only $4\mu$ final states are considered because the $\mathrm{BR}(a^0\rightarrow e^+e^-)$ is vanishingly small in many models.

This analysis is further broken into two separate search regions, which require slightly different approaches due to the different final states and background compositions. One such region searches in the range $1<m_{a^0,Z_d} <15$ GeV, only using the 4$\mu$ final state. The other search region searches the range $15<m_{a^0,Z_d} <60$ GeV, and uses final states containing pairs of electrons and muons, though it has limited sensitivity to the $a^0$ interpretation due to the Yukawa-like couplings of the $a^0$. 

\begin{figure}[htb]
  \centering
  \subfloat[Low Mass]{\includegraphics[width=0.45\textwidth]{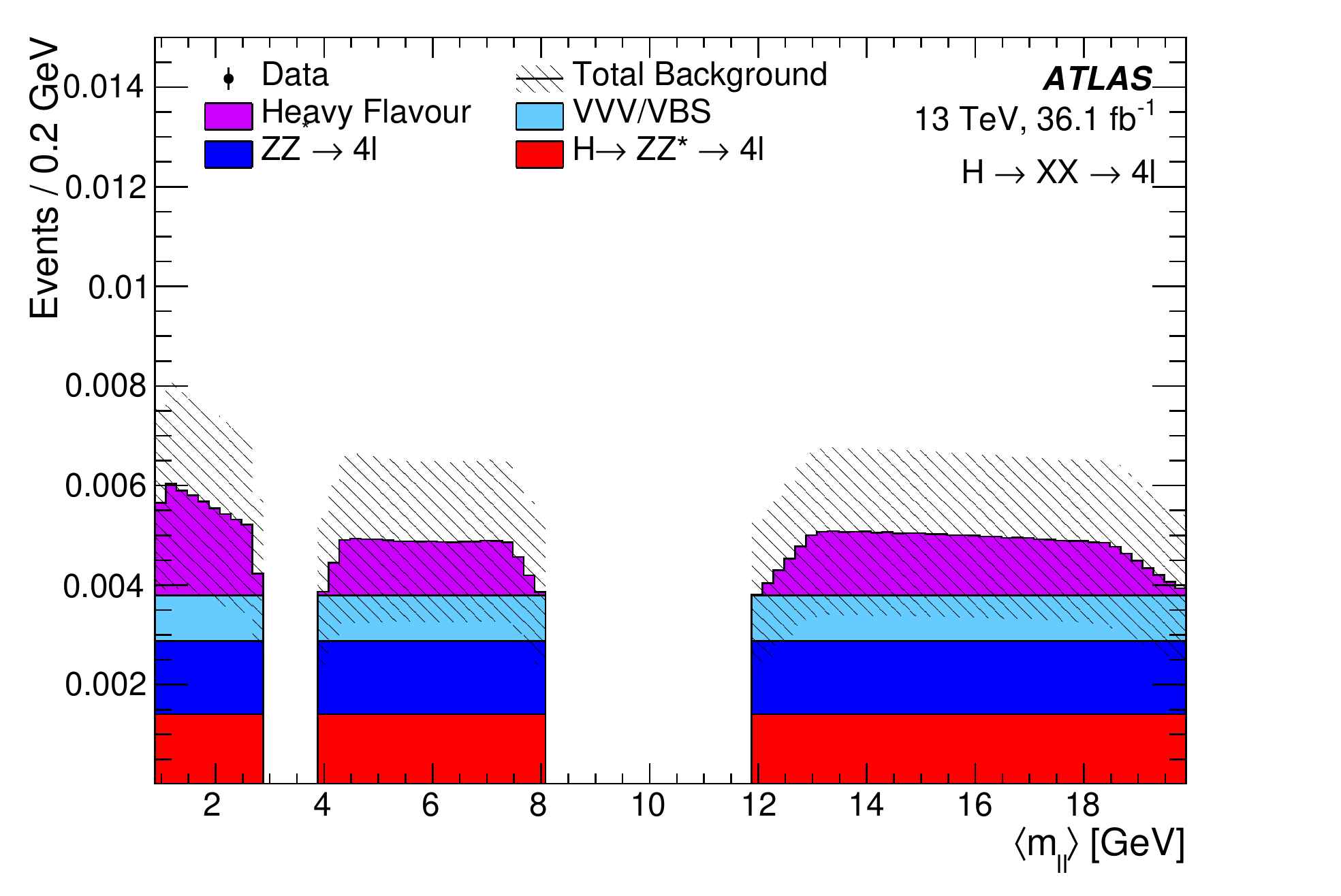}}
  \hspace{0.0333333333\textwidth}
  \subfloat[High Mass]{\includegraphics[width=0.45\textwidth]{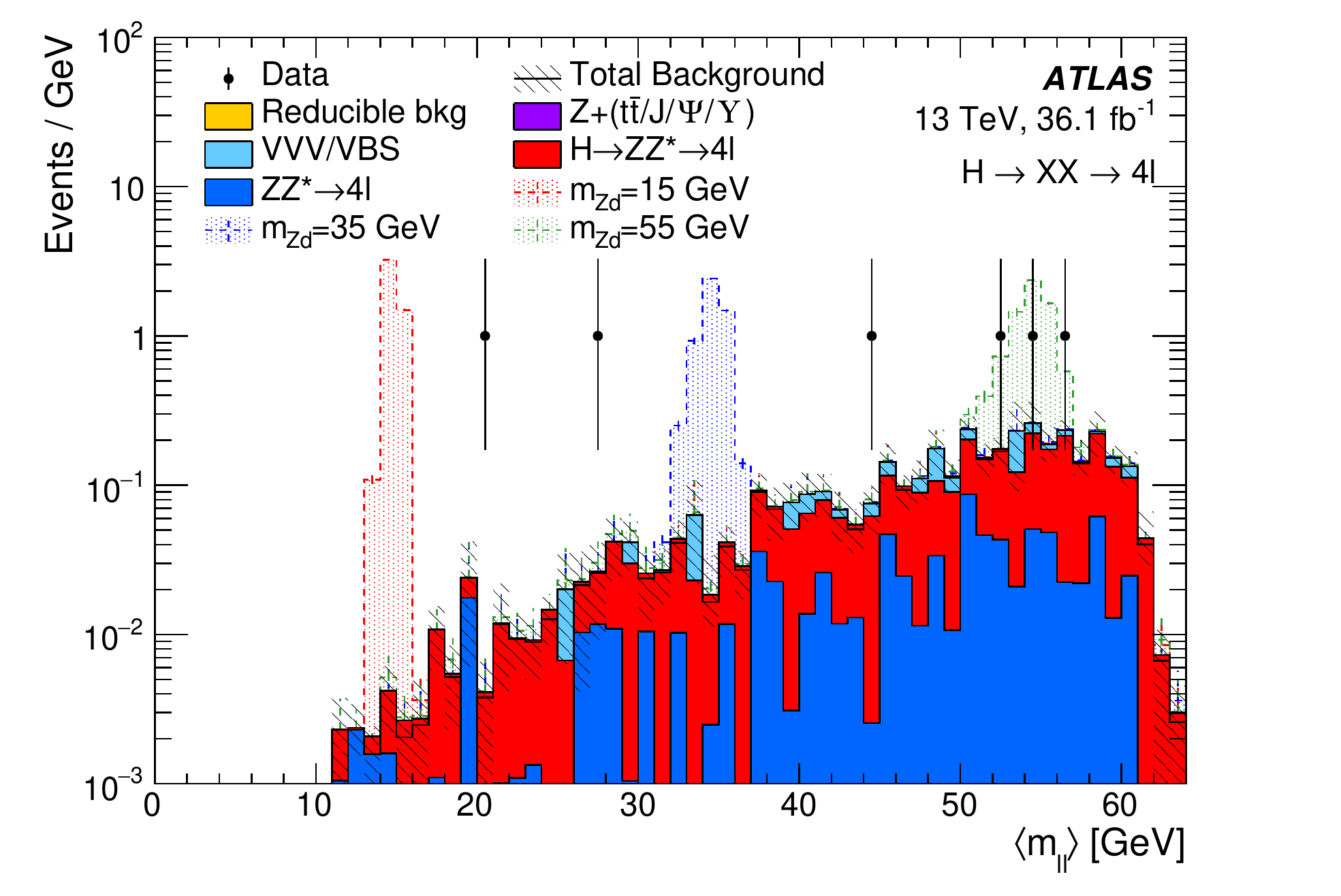}}
  \caption{$\langle m_{\ell\ell}\rangle$ distributions in low (a) and high (b) mass search regions. These distributions are the signal regions used in the fit to obtain the final 95\% $CL_s$ limit~\cite{hxx4l}.}
  \label{fig:hxx4musigreg}
\end{figure}

\begin{figure}[htb]
  \centering
  \subfloat[$a^0$]{\includegraphics[width=0.45\textwidth]{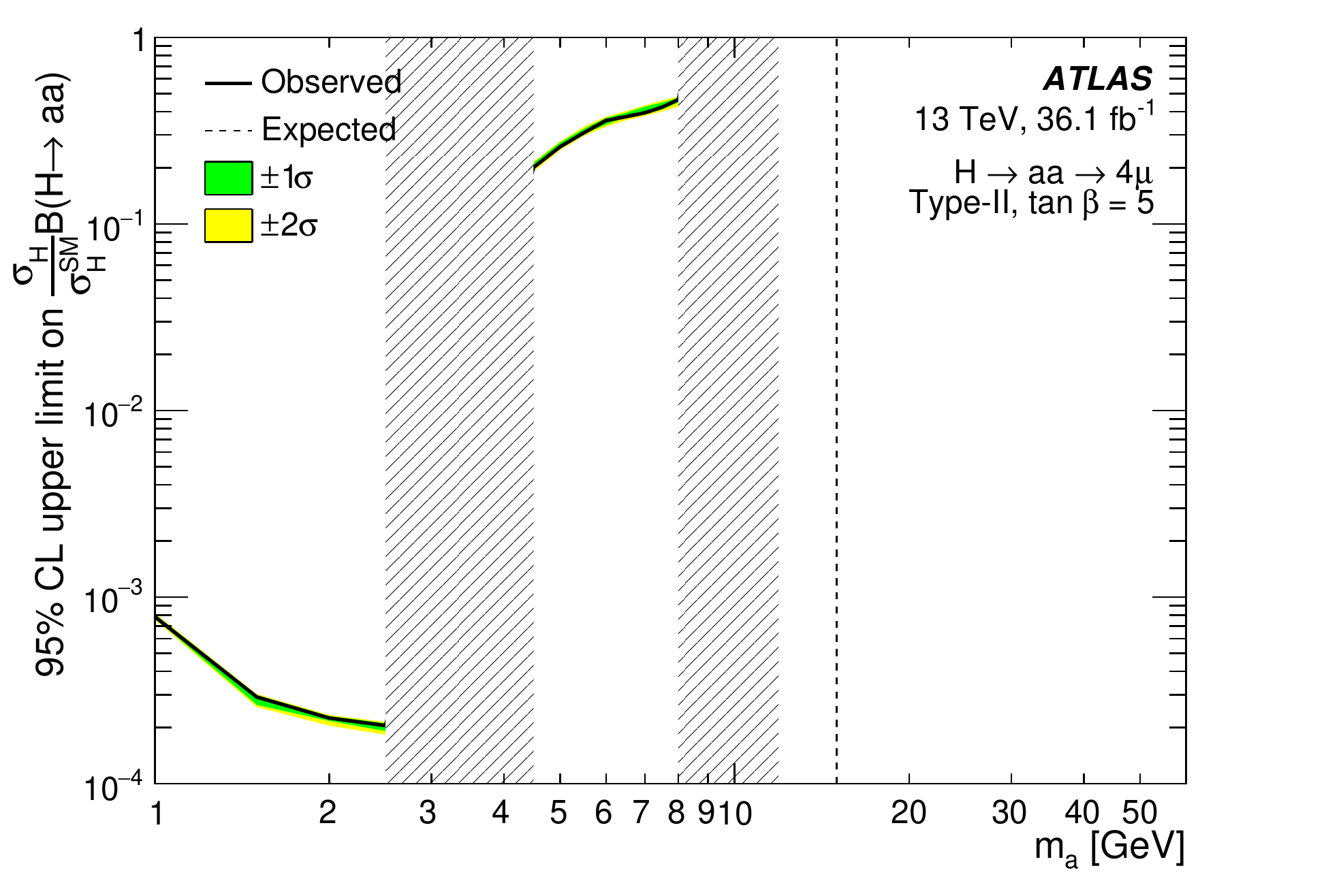}}
  \hspace{0.0333333333\textwidth}
  \subfloat[$Z_d$]{\includegraphics[width=0.45\textwidth]{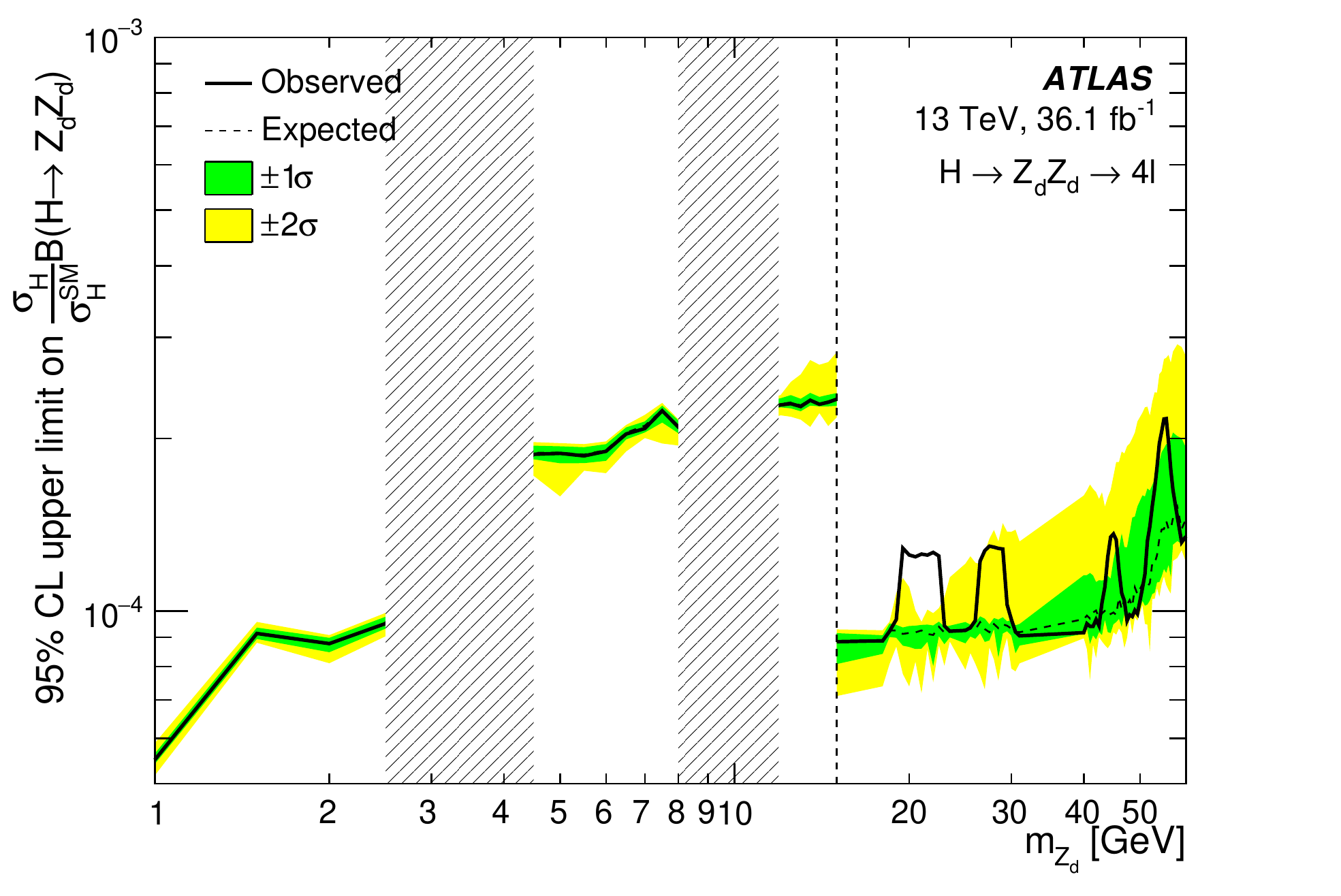}}
  \caption{95\% $CL_s$ limits on $\frac{\sigma_H}{\sigma_H^{\mathrm{SM}}}BR(H\rightarrow a^0a^0)$ (a) and $\frac{\sigma_H}{\sigma_H^{\mathrm{SM}}}BR(H\rightarrow Z_dZ_d)$ (b)~\cite{hxx4l}.}
  \label{fig:hxx4mulimits}
\end{figure}

Lepton quadruplets are formed from pairs of same-flavour opposite-sign isolated leptons. If multiple pairings are possible in an event, pairs are selected which minimise the di-lepton mass difference: $\Delta m=|m_{12}-m_{34}|$. A kinematic selection is then applied, based in part on the invariant masses of the di-lepton pairs and quadruplet, differing only where necessary between the two search regions. The mean di-lepton mass ($\langle m_{\ell\ell}\rangle =(m_{12}+m_{34})/2$) is then used in the fit. Figure~\ref{fig:hxx4musigreg} shows the $\langle m_{\ell\ell}\rangle$ distributions of the SM background and observed data events, for the low and high mass search regions. No (6) events are seen in the low (high) mass range, resulting in the limits shown in Figure~\ref{fig:hxx4mulimits}. 

\subsection{\boldmath $H\rightarrow ZZ_d\rightarrow \ell\ell\ell\ell$}

Hidden dark-sector models predict kinetic mixing between the SM and dark-sector Z bosons, as shown in Figure~\ref{fig:hxxproduction}(c), potentially resulting in an observable signal in the $4\ell$ final state~\cite{hxx4l}. This process can be searched for using an analysis strategy closely resembling that of the search for two light resonances in the $4\ell$ final state, with quadruplets formed from two di-lepton pairs. However, in the event of multiple possible lepton pairings, the di-lepton pairing with a di-lepton mass closest to $m_Z$ is selected. Dark-sector Z bosons are searched for in the range $15<m_{Z_d}<55$ GeV. The dominant backgrounds are $ZZ^*$ and $H\rightarrow ZZ^*$, which are estimated from simulation. A small fake lepton background is estimated using a data-driven method.

\begin{figure}[htb]
  \centering
  \subfloat[Signal Region]{\includegraphics[width=0.45\textwidth]{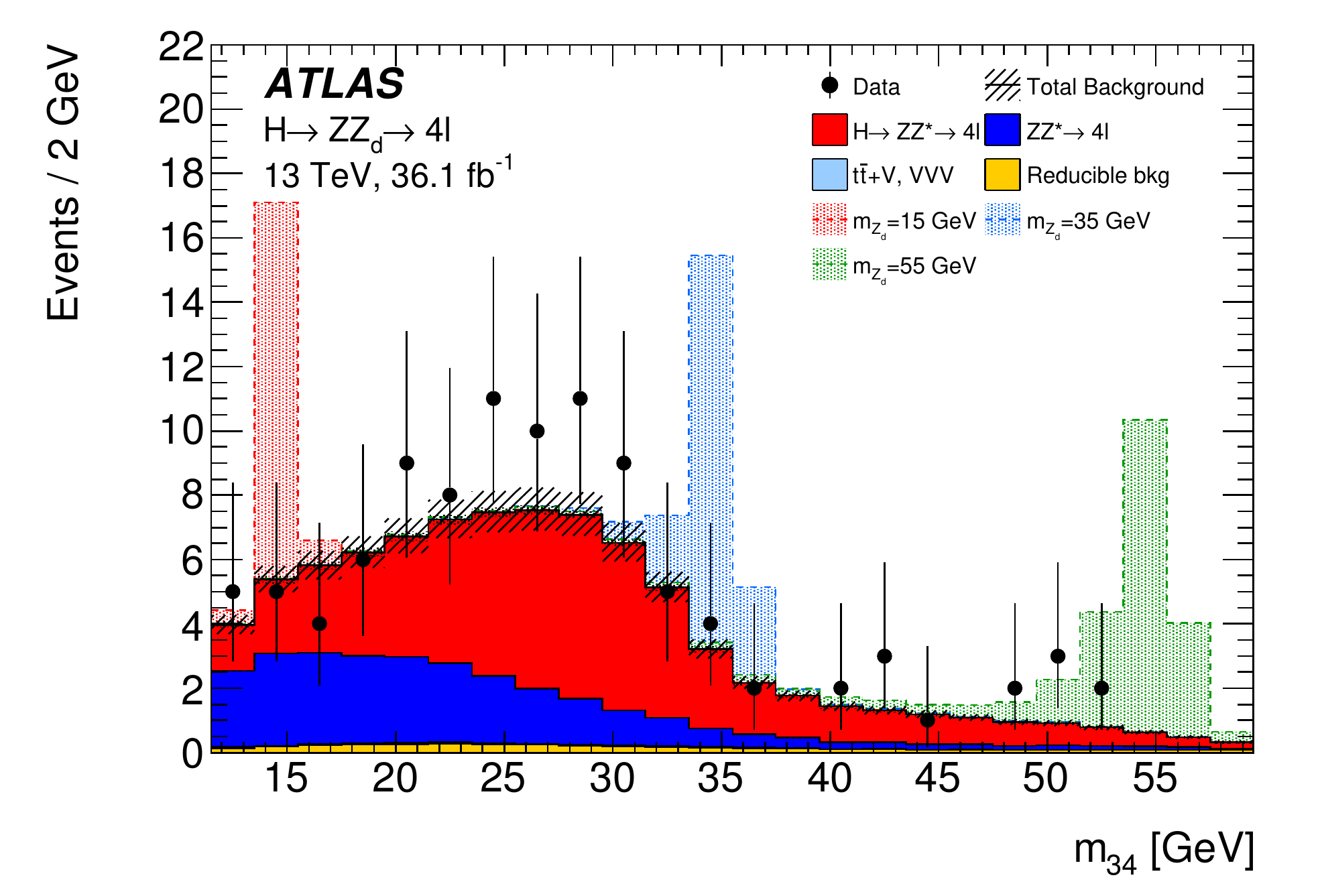}}
  \hspace{0.0333333333\textwidth}
  \subfloat[Expected and observed limits]{\includegraphics[width=0.45\textwidth]{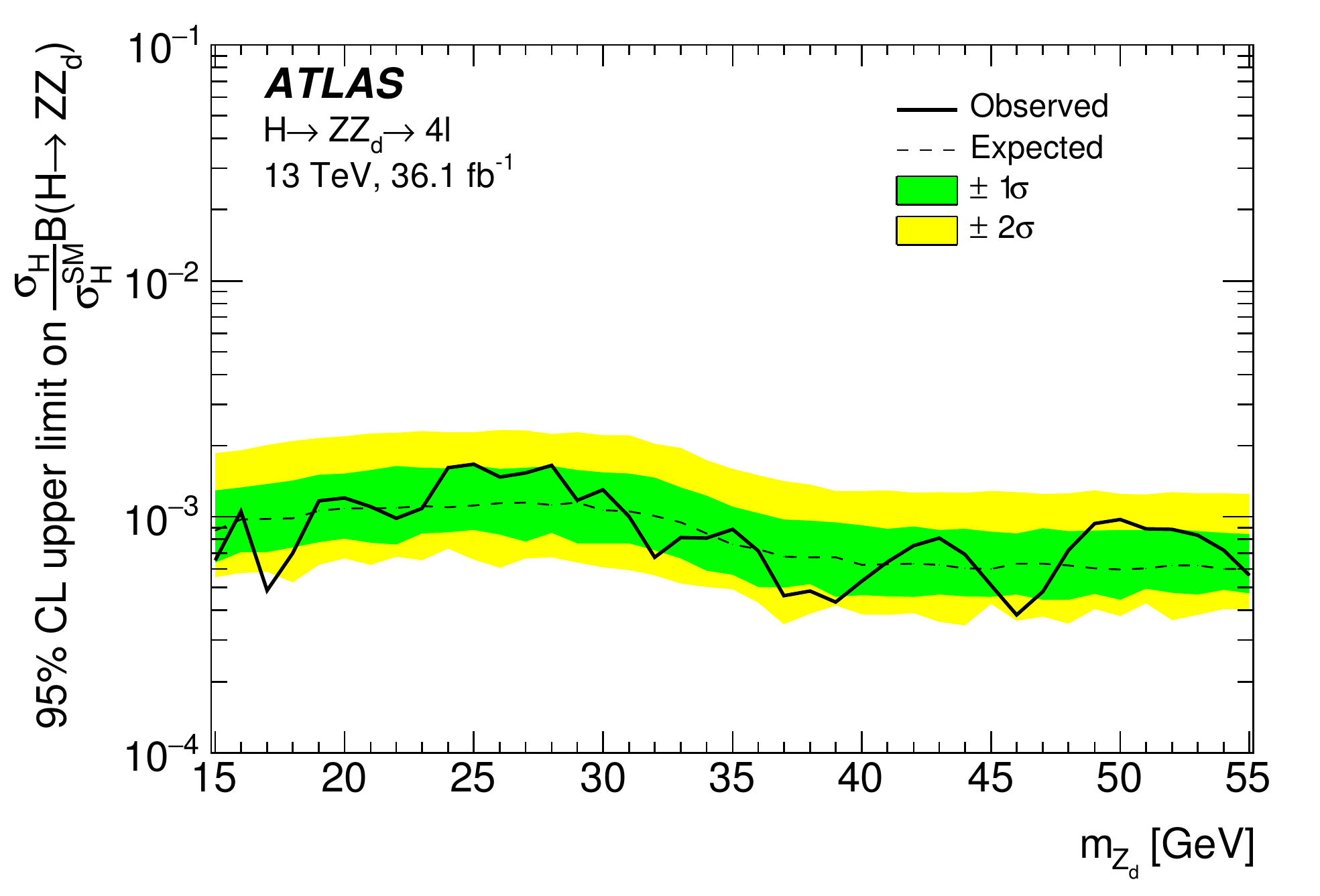}}
  \caption{Invariant mass distribution (a) of the di-lepton pair which is furthest from $m_Z$~\cite{hxx4l}. These distributions are the signal regions used in the fit to obtain the final 95\% $CL_s$ limit (b)~\cite{hxx4l}.}
  \label{fig:zzdresults}
\end{figure}

Events are categorised based on the flavour of the di-lepton system with an invariant mass furthest from $m_Z$: 65 events are observed for di-muons, and 37 for di-electrons. The invariant mass distribution (Figure~\ref{fig:zzdresults}(a)) of this di-lepton pair is fitted to set the limit (Figure~\ref{fig:zzdresults}(b)).

\subsection{\boldmath $H\rightarrow a^0a^0\rightarrow \gamma\gamma jj$}

Searches for $H\rightarrow a^0a^0\rightarrow \gamma\gamma jj$ are sensitive to models in which the fermionic decay modes of the $a^0$ are suppressed~\cite{haagamgamjj}. The search range is $20<m_{a^0}<60$ GeV, with jets being primarily gluon-induced. Due to the large multi-jet backgrounds the vector boson fusion production mode is targeted, though there is significant contribution from the gluon-fusion production mode. After the kinematic selection is applied, the main backgrounds are $\gamma\gamma jj$ and $jjjj$. Independent inversions of the $|m_{jj}-m_{\gamma\gamma}|$ and photon identification requirements is used to estimate the remaining background.

\begin{figure}[htb]
  \centering
  \includegraphics[width=0.45\textwidth]{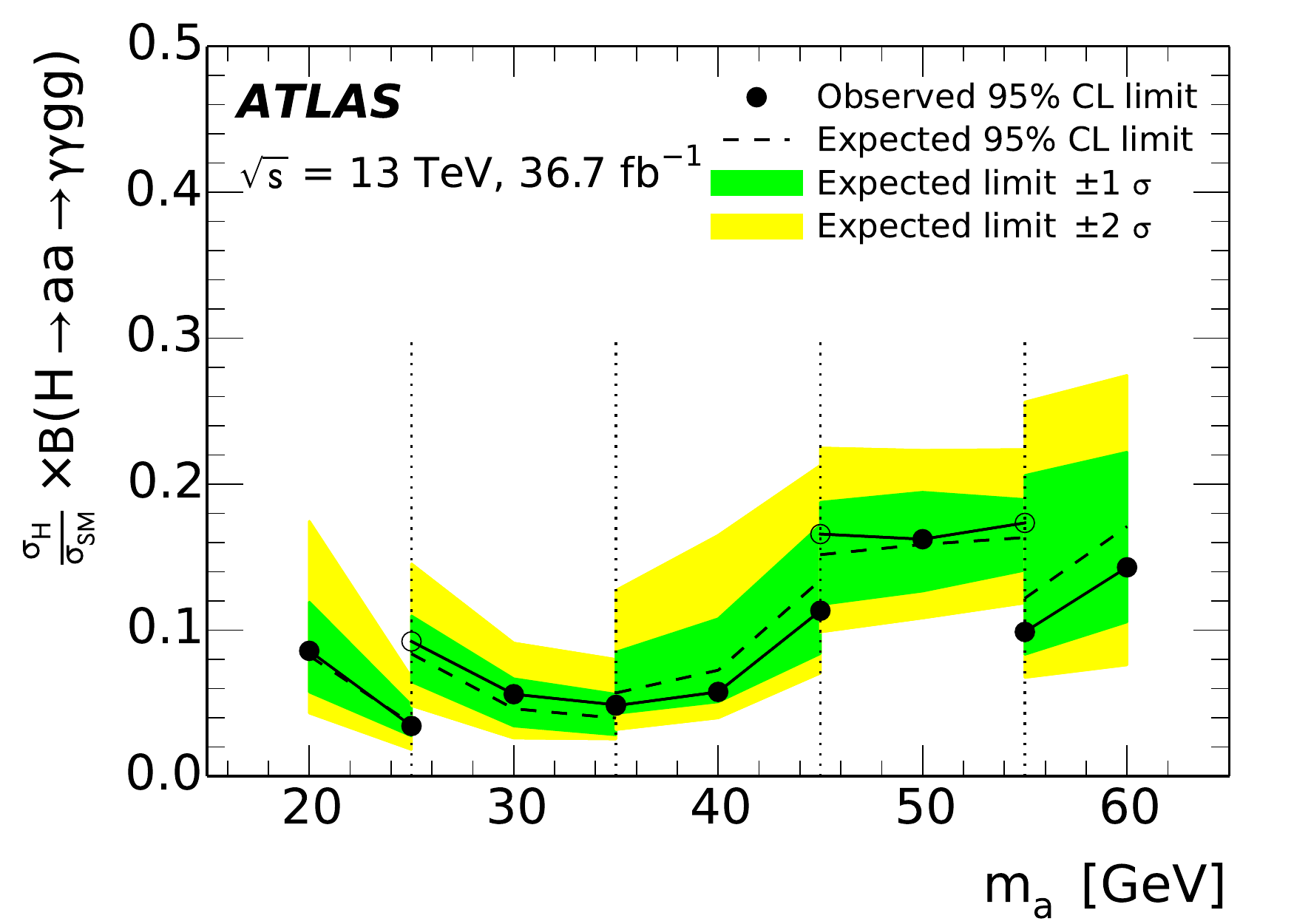}
  \caption{95\% $CL_s$ upper limit on $\frac{\sigma_H}{\sigma_H^{\mathrm{SM}}}BR(H\rightarrow a^0a^0\rightarrow \gamma\gamma gg)$~\cite{haagamgamjj}.}
  \label{fig:gamgamjjlim}
\end{figure}

A fit to five non-exclusive bins in $m_{jj}$ is used to set the limit shown in Figure~\ref{fig:gamgamjjlim}.

\subsection{\boldmath $WH\rightarrow \ell\nu a^0a^0\rightarrow \ell\nu 4b$}

Over a large mass range, the largest branching ratio of the $a^0$ will be to $B$-hadrons~\cite{haalv4b}. However, measuring these is difficult at a hadron collider experiment due to the large multi-jet backgrounds. These backgrounds can be reduced by targeting Higgs bosons produced in association with a leptonically decaying W boson, which also provides an efficient triggering strategy. This search targets the range $20<m_{a^0}<60$ GeV. Eight event categories are defined as: ($n_{jets}$=3,4,5+)x($n_{b-tags}$=2,3,4+), three of which are designated as signal categories: (4j,4b), (4j,3b) and (3j,3b). The dominant background to this search is $t\bar t\ (+jj)$.

\begin{figure}[htb]
  \centering
  \subfloat[$3j3b$]{\includegraphics[width=0.3\textwidth]{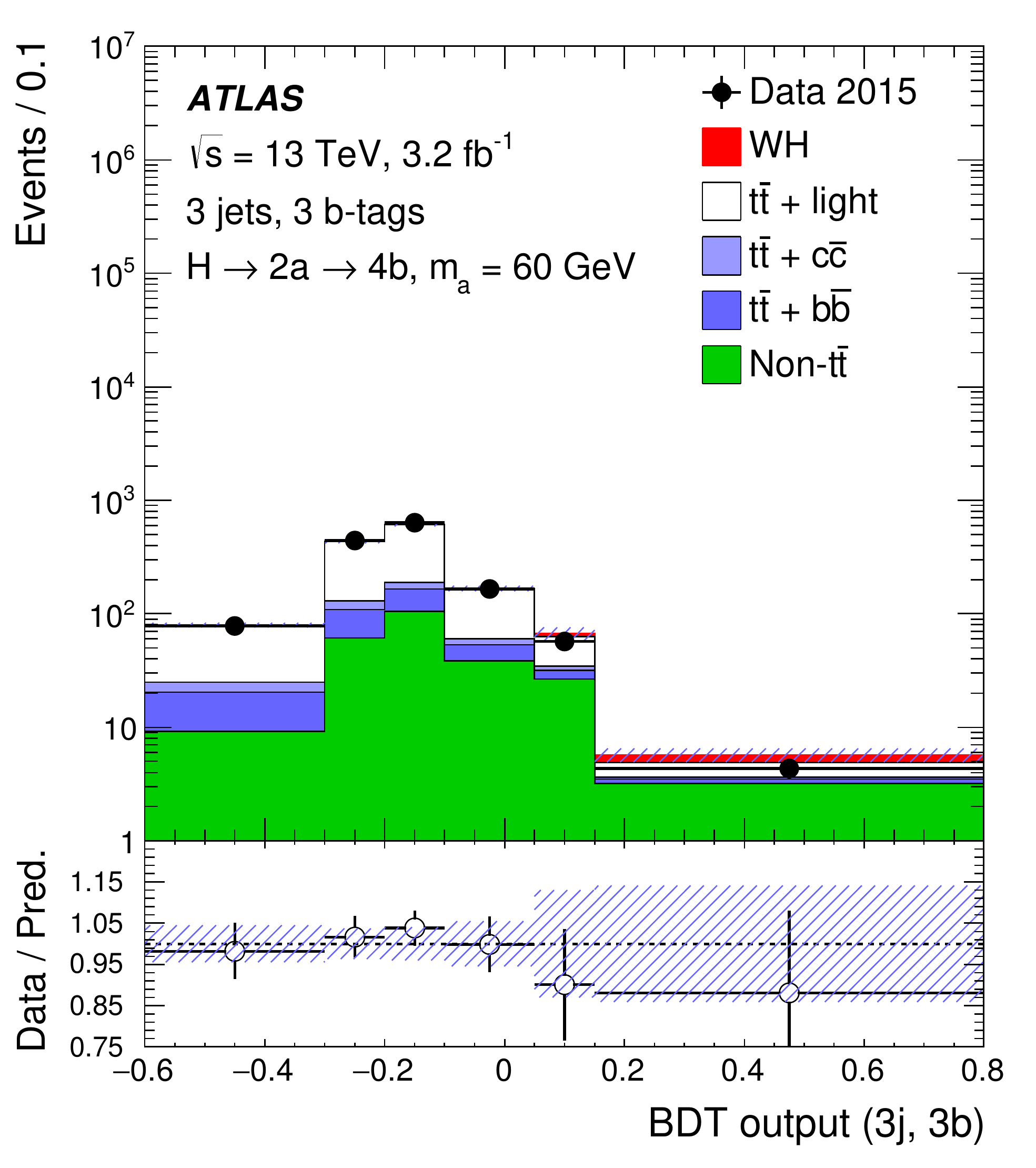}}
  \hspace{0.025\textwidth}
  \subfloat[$4j3b$]{\includegraphics[width=0.3\textwidth]{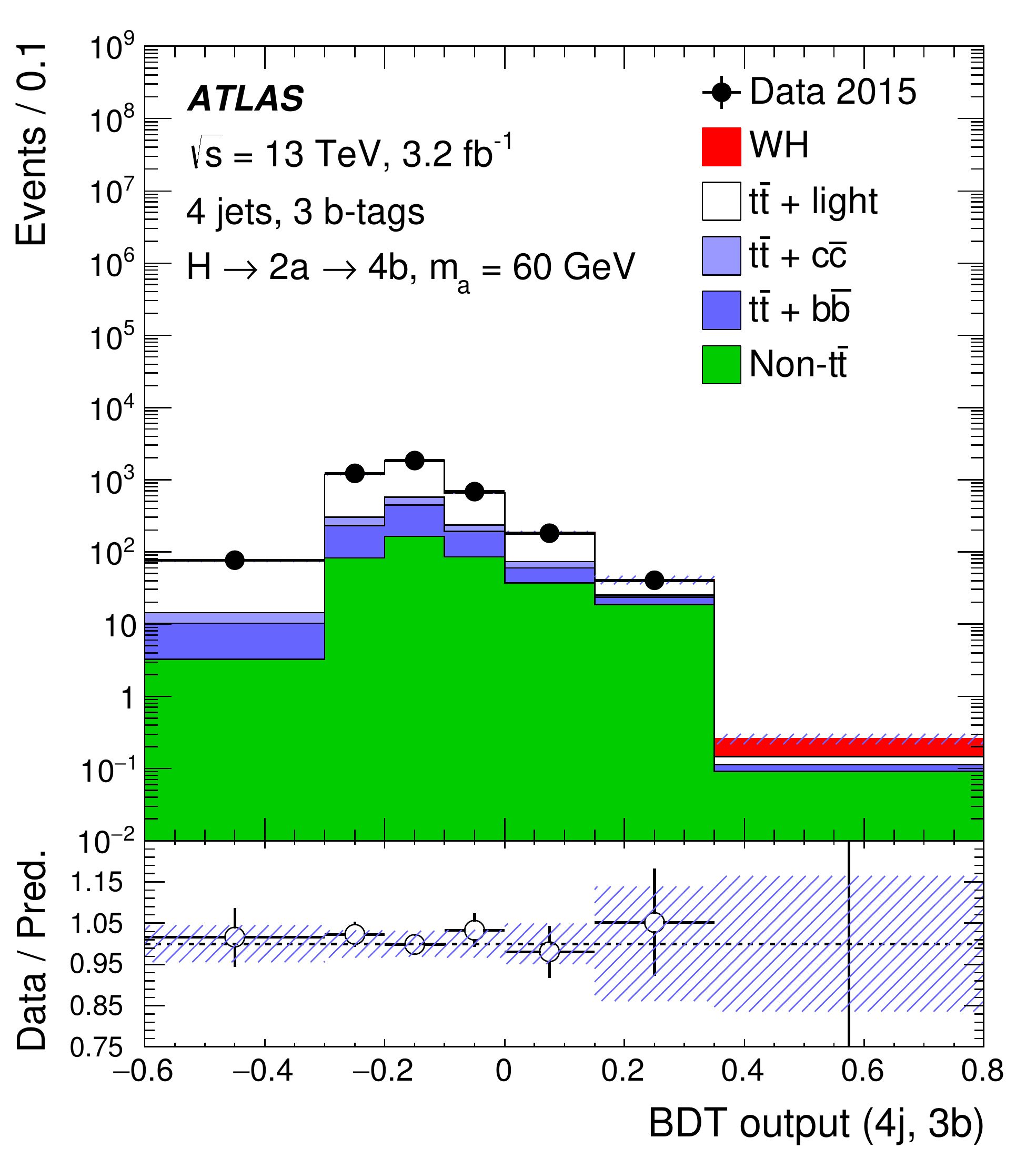}}
  \hspace{0.025\textwidth}
  \subfloat[$4j4b$]{\includegraphics[width=0.3\textwidth]{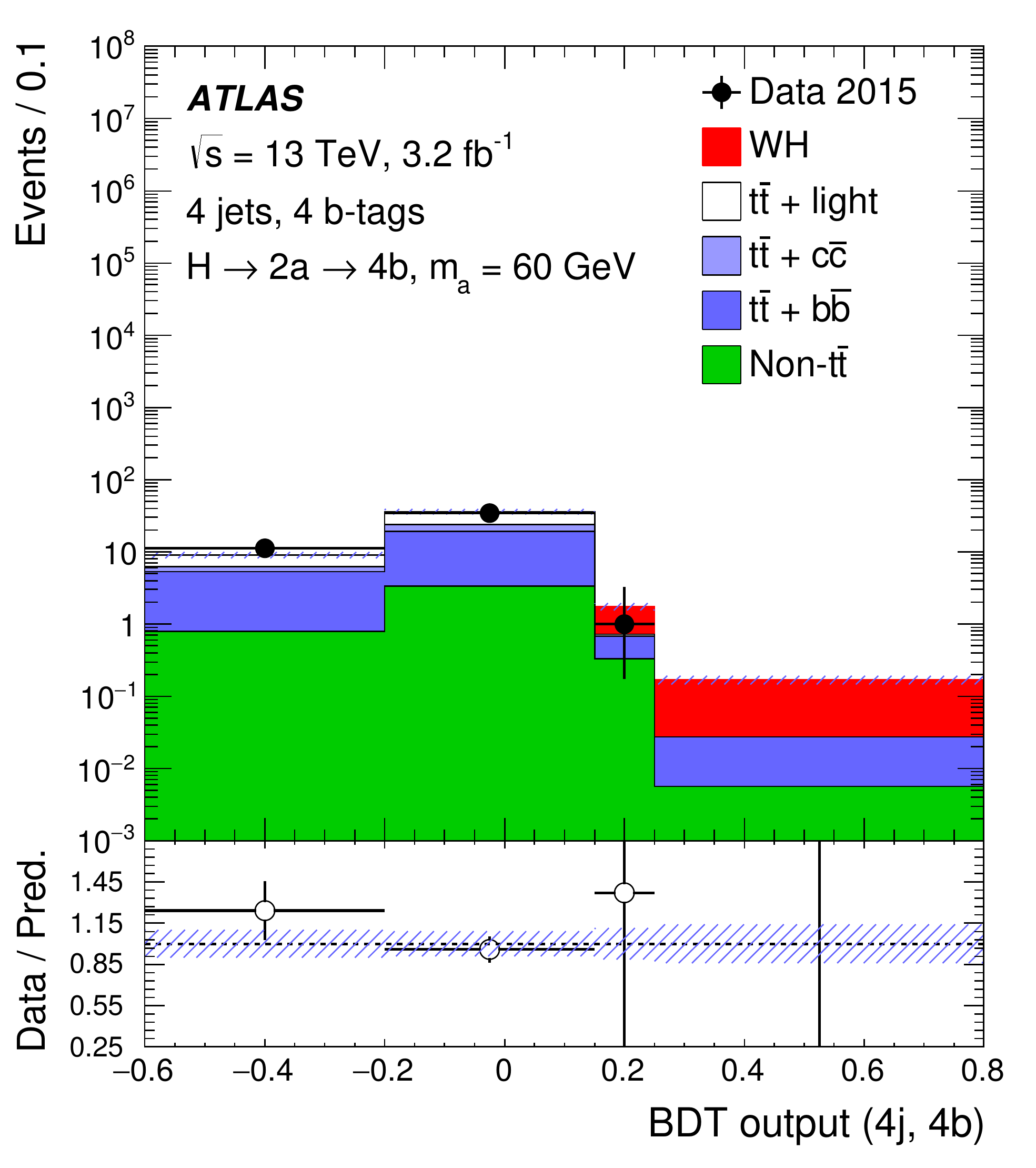}}
  \caption{BDT output variables, used as input to the fit for the three signal regions in the $WH\rightarrow \ell\nu a^0a^0\rightarrow \ell\nu 4b$ search~\cite{haalv4b}.}
  \label{fig:bdts}
\end{figure}

\begin{figure}[htb]
  \centering
  \includegraphics[width=0.45\textwidth]{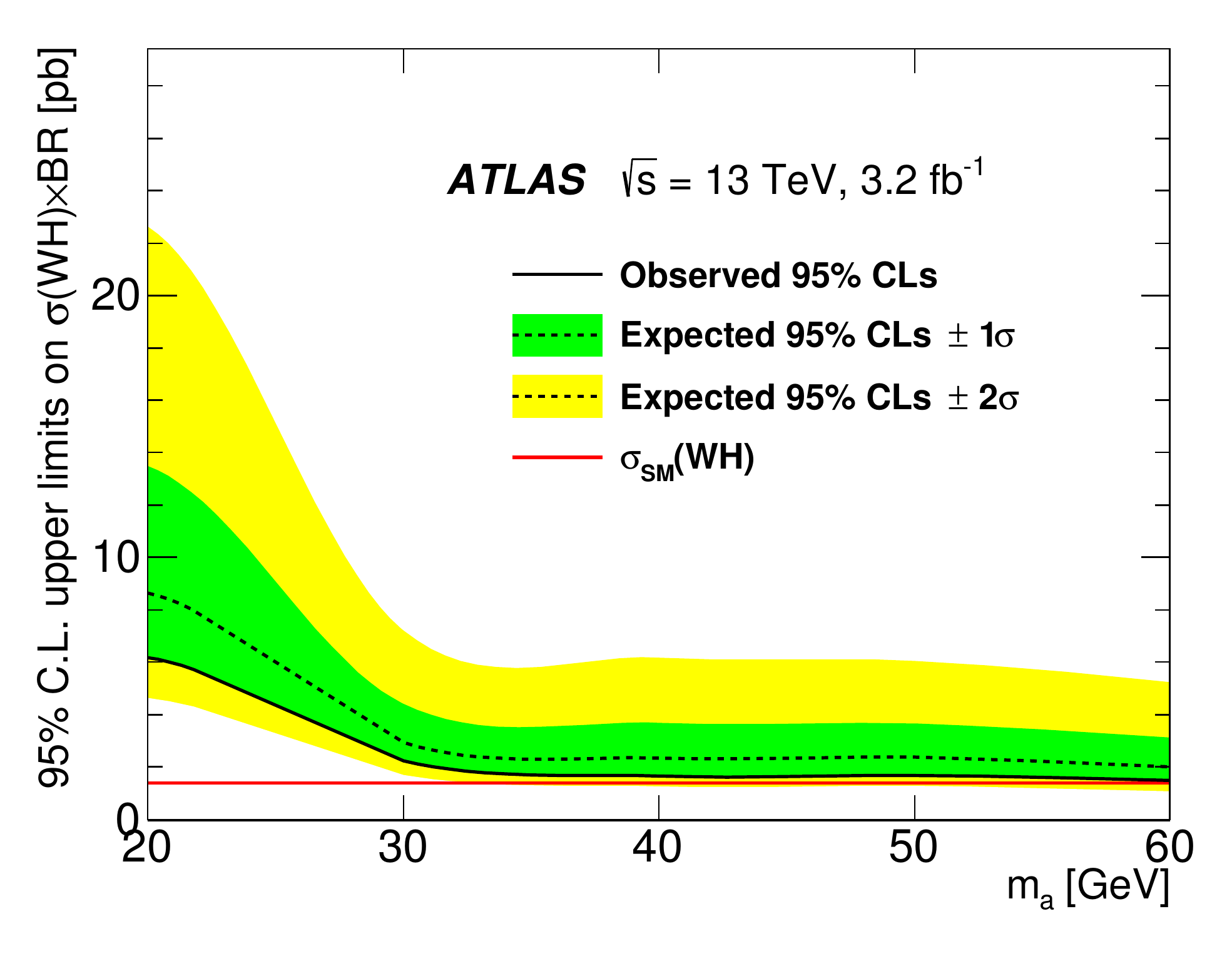}
  \caption{95\% $CL_s$ upper limit on $\sigma(WH)\times BR$~\cite{haalv4b}.}
  \label{fig:lv4blim}
\end{figure}

Nine kinematic variables are used as inputs to a boosted decision tree (BDT), the output (Figure~\ref{fig:bdts}) of which forms the final discriminant for the three signal regions. The five background regions use the sum of the transverse hadronic energy as the final discriminant. A simultaneous fit across all eight event categories is used to set the final limit, shown in Figure~\ref{fig:lv4blim}.

\section{Summary}

Various searches were presented for rare and BSM decays of the Higgs boson with the ATLAS detector at the LHC. While no searches to date have resulted in an observation of a significant excess, they are all statistically limited, and so should improve with the square root of the delivered integrated luminosity. This is especially promising looking to the High-Luminosity LHC, where for example, the limit for the $H\rightarrow J/\psi\gamma$ search is set to improve to $\sim 15\times \sigma_{SM}$~\cite{jpsiprospects}.




\begin{thebibliography}{99}

\bibitem{couplings}
ATLAS Collaboration, \emph{Measurements of the Higgs boson production and decay rates and coupling strengths using $pp$ collision data at $\sqrt{s}$ = 7 and 8 TeV in the ATLAS experiment}, \href{https://link.springer.com/article/10.1140\%2Fepjc\%2Fs10052-015-3769-y}{Eur. Phys. J. C {\bf 76} (2016) 6}, arXiv: \href{https://arxiv.org/abs/1507.04548}{1507.04548 [hep-ex]}.

\bibitem{masscouplingscms}
CMS Collaboration, \emph{Precise determination of the mass of the Higgs boson and tests of compatibility of its couplings with the standard model predictions using proton collisions at 7 and 8 TeV}, \href{https://link.springer.com/article/10.1140\%2Fepjc\%2Fs10052-015-3351-7}{Eur. Phys. J. C {\bf 75} (2015) 212}, arXiv: \href{https://arxiv.org/abs/1412.8662}{1412.8662 [hep-ex]}.

\bibitem{masscouplingszcms}
CMS Collaboration, \emph{Study of the Mass and Spin-Parity of the Higgs Boson Candidate Via Its Decays to $Z$ Boson Pairs}, \href{https://journals.aps.org/prl/abstract/10.1103/PhysRevLett.110.081803}{Phys. Rev. Lett. {\bf 110} (2013) 081803}, arXiv: \href{https://arxiv.org/abs/1212.6639}{1212.6639 [hep-ex]}.

\bibitem{spin0}
ATLAS Collaboration, \emph{Evidence for the spin-0 nature of the Higgs boson using ATLAS data}, \href{https://www.sciencedirect.com/science/article/pii/S0370269313006527?via\%3Dihub}{Phys. Lett. B {\bf726} (2013) 120}, arXiv: \href{https://arxiv.org/abs/1307.1432}{1307.1432 [hep-ex]}.

\bibitem{anomolouscouplingscms}
CMS Collaboration, \emph{Constraints on the spin-parity and anomalous $HVV$ couplings of the Higgs boson in proton collisions at 7 and 8 TeV}, \href{https://journals.aps.org/prd/abstract/10.1103/PhysRevD.92.012004}{Phys. Rev. D {\bf 92} (2015) 012004}, arXiv: \href{https://arxiv.org/abs/1411.3441}{1411.3441 [hep-ex]}.

\bibitem{couplingsatlasandcms}
ATLAS and CMS Collaborations, \emph{Measurements of the Higgs boson production and decay rates and constraints on its couplings from a combined ATLAS and CMS analysis of the LHC $pp$ collision data at $\sqrt{s}$ = 7 and 8 TeV}, \href{https://link.springer.com/article/10.1007\%2FJHEP08\%282016\%29045}{JHEP {\bf 08} (2016) 045}, arXiv: \href{https://arxiv.org/abs/1606.02266}{1606.02266 [hep-ex]}.

\bibitem{massatlasandcms}
ATLAS and CMS Collaborations, \emph{Combined Measurement of the Higgs Boson Mass in pp Collisions at $sqrt{s}$ = 7 and 8 TeV with the ATLAS and CMS Experiments}, \href{https://journals.aps.org/prl/abstract/10.1103/PhysRevLett.114.191803}{Phys. Rev. Lett. {\bf 114} (2015) 191803}, arXiv: \href{https://arxiv.org/abs/1503.07589}{1503.07589 [hep-ex]}.

\bibitem{atlas}
ATLAS Collaboration, \emph{The ATLAS Experiment at the CERN Large Hadron Collider}, \href{http://iopscience.iop.org/article/10.1088/1748-0221/3/08/S08003/meta}{JINST {\bf 3} (2008) S08003}.

\bibitem{lhc}
L. Evans (ed.), P. Bryant (ed.) (CERN), \emph{LHC Machine}, \href{http://iopscience.iop.org/article/10.1088/1748-0221/3/08/S08001/meta}{JINST {\bf 3} (2008) S08001}.

\bibitem{CLs1}
G. Cowan, K. Cranmer, E. Gross, and O. Vitells, \emph{Asymptotic formulae for likelihood-based tests of new physics}, \href{https://link.springer.com/article/10.1140\%2Fepjc\%2Fs10052-011-1554-0}{Eur. Phys. J. C {\bf 71} (2011) 1554}, arXiv: \href{https://arxiv.org/abs/1007.1727}{1007.1727 [physics.data-an]}, Erratum: \href{https://link.springer.com/article/10.1140\%2Fepjc\%2Fs10052-013-2501-z}{{Eur. Phys. J. C {\bf 73} (2013) 2501}}.

\bibitem{CLs2}
A. L. Read, \emph{Presentation of search results: the $CL_s$ technique}, \href{http://iopscience.iop.org/article/10.1088/0954-3899/28/10/313/meta}{J. Phys. G {\bf 28} (2002) 2693}.

\bibitem{Hcc}
ATLAS Collaboration, \emph{Search for the Decay of the Higgs Boson to Charm Quarks with the ATLAS Experiment}, \href{https://journals.aps.org/prl/abstract/10.1103/PhysRevLett.120.211802}{Phys. Rev. Lett. {\bf 120} (2018) 211802}, arXiv: \href{https://arxiv.org/abs/1802.04329}{1802.04329 [hep-ex]}.

\bibitem{Hmumu}
ATLAS Collaboration, \emph{Search for the Dimuon Decay of the Higgs Boson in $pp$ Collisions at $\sqrt{s}$ = 13 TeV with the ATLAS Detector}, \href{https://journals.aps.org/prl/abstract/10.1103/PhysRevLett.119.051802}{Phys. Rev. Lett. {\bf 119}, 051802 (2017)}, arXiv: \href{https://arxiv.org/abs/1705.04582}{1705.04582 [hep-ex]}.

\bibitem{Hinv}
ATLAS Collaboration, \emph{Search for an invisibly decaying Higgs boson or dark matter candidates produced in association with a $Z$ boson in $pp$ collisions at $\sqrt{s}$ = 13 TeV with the ATLAS detector}, \href{https://www.sciencedirect.com/science/article/pii/S0370269317309413?via\%3Dihub}{PLB {\bf 776} (2017), 318}, arXiv: \href{https://arxiv.org/abs/1708.09624}{1708.09624 [hep-ex]}.

\bibitem{HZgam}
ATLAS Collaboration, \emph{Searches for the $Z\gamma$ decay mode of the Higgs boson and for new high-mass resonances in $pp$ collisions at $\sqrt{s}$=13 TeV with the ATLAS detector}, \href{https://link.springer.com/article/10.1007\%2FJHEP10\%282017\%29112}{JHEP {\bf 10} (2017) 112}, arXiv: \href{https://arxiv.org/abs/1708.00212}{1708.00212 [hep-ex]}.

\bibitem{hmgam1}
ATLAS Collaboration, \emph{Search for exclusive Higgs and $Z$ boson decays to $\phi\gamma$ and $\rho\gamma$ with the ATLAS detector}, \href{https://link.springer.com/article/10.1007\%2FJHEP07\%282018\%29127}{JHEP {\bf 07} (2018) 127}, arXiv: \href{https://arxiv.org/abs/1712.02758}{1712.02758}.
  
\bibitem{hmgam2}
ATLAS Collaboration, \emph{Search for Higgs and $Z$ Boson Decays to $J/\psi\gamma$ and $\Upsilon(nS)\gamma$ with the ATLAS Detector}, \href{https://journals.aps.org/prl/abstract/10.1103/PhysRevLett.114.121801}{Phys. Rev. Lett. {\bf 114}, 121801 (2015)}, arXiv: \href{https://arxiv.org/abs/1501.03276}{1501.03276 [hep-ex]}.

\bibitem{hmgam3}
ATLAS Collaboration, \emph{Search for Higgs and $Z$ Boson Decays to $\phi\gamma$ with the ATLAS Detector}, \href{}{Phys. Rev. Lett. {\bf 117}, 111802 (2016)}, arXiv: \href{https://arxiv.org/abs/1607.03400}{1607.03400 [hep-ex]}.

\bibitem{HMgamCalc1}
M. Koenig, M. Neubert, \emph{Exclusive Radiative Higgs Decays as Probes of Light-Quark Yukawa Couplings}, \href{https://link.springer.com/article/10.1007\%2FJHEP08\%282015\%29012}{JHEP {\bf 1508} (2015) 012}, arXiv: \href{https://arxiv.org/abs/1505.03870}{1505.03870 [hep-ph]}.

\bibitem{HMgamCalc2}
G. T. Bodwin, H. S. Chung, J.-H. Ee, J. Lee, F. Petriello, \emph{Relativistic corrections to Higgs boson decays to quarkonia}, \href{https://journals.aps.org/prd/abstract/10.1103/PhysRevD.90.113010}{Phys. Rev. D {\bf 90}, 113010 (2014)}, arXiv: \href{https://arxiv.org/abs/1407.6695}{1407.6695 [hep-ph]}.

\bibitem{20pc}
  C. Delaunay, T. Golling, G. Perez, and Y. Soreq, \emph{Enhanced Higgs boson coupling to charm pairs}, \href{https://journals.aps.org/prd/abstract/10.1103/PhysRevD.89.033014}{Phys. Rev. D {\bf 89} (2014) 033014}, arXiv: \href{https://arxiv.org/abs/1310.7029}{1310.7029 [hep-ph]}.

\bibitem{nmssm}
G.F. Giudice and A. Masiero, \emph{A natural solution to the μ problem in supergravity theories}, \href{https://www.sciencedirect.com/science/article/pii/0370269388916139?via\%3Dihub}{Phys. Lett. B {\bf 206} (1988) 480}.

\bibitem{hahm}
D. Curtin, R. Essig, S. Gori and J. Shelton, \emph{Illuminating dark photons with high-energy colliders}, \href{https://link.springer.com/article/10.1007\%2FJHEP02\%282015\%29157}{JHEP {\bf 02} (2015) 157}, arXiv: \href{https://arxiv.org/abs/1412.0018}{1412.0018 [hep-ph]}.

\bibitem{mumutautau}
ATLAS Collaboration, \emph{Search for Higgs bosons decaying to $aa$ in the $\mu\mu\tau\tau$ final state in $pp$ collisions at $\sqrt{s}$ = 8 TeV with the ATLAS experiment}, CERN-PH-EP-2015-057, arXiv: \href{https://arxiv.org/abs/1505.01609}{1505.01609 [hep-ex]}.

\bibitem{gamgamgamgam}
ATLAS Collaboration, \emph{Search for new phenomena in events with at least three photons collected in $pp$ collisions at $\sqrt{s}$ = 8 TeV with the ATLAS detector}, \href{https://link.springer.com/article/10.1140\%2Fepjc\%2Fs10052-016-4034-8}{Eur. Phys. J. C {\bf 76} (2016)}, arXiv: \href{https://arxiv.org/abs/1509.05051}{1509.05051 [hep-ex]}.

\bibitem{hxx4l}
  ATLAS Collaboration, \emph{Search for Higgs boson decays to beyond-the-Standard-Model light bosons in four-lepton events with the ATLAS detector at $\sqrt{s}$=13 TeV}, \href{https://link.springer.com/article/10.1007\%2FJHEP06\%282018\%29166}{JHEP {\bf 06} (2018) 166}, arXiv: \href{https://arxiv.org/abs/1802.03388}{arXiv:1802.03388 [hep-ex]}.

\bibitem{Leney:2282407}
K. Leney on behalf of the ATLAS Collaboration, \emph{Search for non-standard and rare decays of the Higgs boson with the ATLAS detector}, \href{https://cds.cern.ch/record/2282407}{ATL-PHYS-SLIDE-2017-719}.

\bibitem{a0brs}
D. Curtin, R. Essig, S. Gori, P. Jaiswal, A. Katz et al., \emph{Exotic decays of the 125 GeV Higgs boson}, \href{https://journals.aps.org/prd/abstract/10.1103/PhysRevD.90.075004}{Phys. Rev. D {\bf 90} (2014) 075004}, arXiv: \href{https://arxiv.org/abs/1312.4992}{1312.4992 [hep-ph]}.

\bibitem{haagamgamjj}
ATLAS Collaboration, \emph{Search for Higgs boson decays into pairs of light (pseudo)scalar particles in the $\gamma\gamma jj$ final state in $pp$ collisions at $\sqrt{s}$=13 TeV with the ATLAS detector}, \href{https://www.sciencedirect.com/science/article/pii/S0370269318304593?via\%3Dihub}{Phys. Lett. B {\bf 782} (2018) 750}, arXiv: \href{https://arxiv.org/abs/1803.11145}{1803.11145 [hep-ex]}.

\bibitem{haalv4b}
ATLAS Collaboration, \emph{Search for the Higgs boson produced in association with a $W$ boson and decaying to four $b$-quarks via two spin-zero particles in $pp$ collisions at 13 TeV with the ATLAS detector}, \href{https://link.springer.com/article/10.1140\%2Fepjc\%2Fs10052-016-4418-9}{Eur. Phys. J. C {\bf 76} (2016) 605}, arXiv: \href{https://arxiv.org/abs/1606.08391}{1606.08391 [hep-ex]}.

\bibitem{jpsiprospects}
ATLAS Collaboration, \emph{Search for the Standard Model Higgs and $Z$ Boson decays to $J/\psi\,\gamma$: HL-LHC projections}, \href{https://cds.cern.ch/record/2054550}{ATL-PHYS-PUB-2015-043} (2015).



\end{thebibliography}
\end{document}